\documentclass[letterpaper, 10 pt, conference]{ieeeconf}  

\usepackage{graphicx} 
\usepackage{epstopdf} 
\usepackage{psfrag} 
\usepackage{amsmath} 
\usepackage{mathtools}
\usepackage[dvipsnames]{xcolor} 
\allowdisplaybreaks 
\usepackage{amssymb}  
\usepackage{amsthm,thmtools,dsfont} 
\usepackage{color} 
\usepackage{tabu} 
\usepackage{tabularx}

\usepackage{cite} 
\usepackage{microtype,enumerate,url} 
\usepackage{BOONDOX-ds} 
\usepackage{tikz} 
\usepackage[fulladjust]{marginnote} 
\usepackage{pgfplotstable} 
\usepackage{pgfplots} 
\usepackage{xstring} 
\usepackage{booktabs} 
\usepackage{colortbl} 
\usepackage[americanresistors,americaninductors]{circuitikz} 
\usetikzlibrary{arrows,patterns} 
\usepackage{caption} 
\usepackage{mathabx} 
\usepackage{algorithm} 
\usepackage{algpseudocode} 
\usepgfplotslibrary{fillbetween} 

\makeatletter
\let\NAT@parse\undefined
\makeatother
\usepackage[hidelinks]{hyperref}
\hypersetup{
    colorlinks=true,
    linkcolor=.,
    linkbordercolor=NavyBlue,
    citebordercolor=OliveGreen,
    filecolor=cyan,      
    urlcolor=OliveGreen,
    citecolor=NavyBlue,
    pdfpagemode=FullScreen,
    }
\usepackage[capitalise,nameinlink,sort]{cleveref}
\let\labelindent\relax
\usepackage{enumitem} 

\usepackage{fontawesome}

\usepackage{pgf, tkz-berge}

\definecolor{Gray}{gray}{0.9}

\newcommand{\mc}[1] {\mathcal{#1}}

\renewcommand{\theenumi}{(\roman{enumi})}

\makeatletter
\def\@IEEEsectpunct{.\ \,}
\def\paragraph{\@startsection{paragraph}{4}{\z@}{1.5ex plus 1.5ex minus 0.5ex}%
{0ex}{\normalfont\normalsize\itshape}}
\makeatother

\renewcommand*\descriptionlabel[1]{\hspace\labelsep\normalfont#1}

\declaretheorem[style=definition,numbered=no]{standing assumption}

\renewcommand\thmcontinues[1]{continued}

\newcommand{\beq}{\begin{equation}}
\newcommand{\eeq}{\end{equation}}
\newcommand {\bseq}{\begin{subequations}}
\newcommand {\eseq}{\end{subequations}}

\def\eblue{\protect\normalcolor}

\ctikzset{
    resistors/scale=0.7,
    capacitors/scale=0.7,
    diodes/scale=0.6,
    inductors/coils=6,
    blocks/scale=0.6
}

\usepackage[acronym]{glossaries}
\makeglossaries

\newacronym{LQT}{LQT}{\textbf{L}inear \textbf{Q}uadratic \textbf{T}racking}
\newacronym{LTI}{LTI}{\textbf{L}inear \textbf{T}ime-\textbf{I}nvariant}
\newacronym{LQT-Split}{LQT-Split}{\textbf{L}inear \textbf{Q}uadratic \textbf{T}racking with Davis-Yin \textbf{Split}ting}
\newacronym{POCS}{POCS}{\textbf{P}rojection \textbf{O}nto \textbf{C}onvex \textbf{S}ets}
\newacronym{LQR}{LQR}{\textbf{L}inear \textbf{Q}uadratic \textbf{R}egulator}
\newacronym{PID}{PID}{\textbf{P}roportional- \textbf{I}ntegral- \textbf{D}erivative}
\newacronym{DYS}{DYS}{\textbf{D}avis- \textbf{Y}in \textbf{S}plitting}
\newacronym{QP}{QP}{\textbf{Q}uadratic  \textbf{P}rogramming}
\newacronym{DeePC}{DeePC}{\textbf{D}ata-\textbf{E}nabled \textbf{P}redictive  \textbf{C}ontrol}
\newacronym{LQT-Split-as-a-Pro}{LQT-Split-as-a-Pro}{\textbf{L}inear \textbf{Q}uadratic \textbf{T}racking via operator \textbf{Split}ting \text{a}nd \textbf{s}calable \textbf{a}lternating \textbf{Pro}jection}
\newacronym{Split-as-a-Pro}{Split-as-a-Pro}{operator \textbf{Split}ting \text{a}nd \textbf{s}calable \textbf{a}lternating \textbf{Pro}jection}
\newacronym{CCP}{CCP}{\textbf{C}onvex, \textbf{C}losed, and \textbf{P}roper}
\newacronym{ITOS}{ITOS}{\textbf{I}nexact \textbf{T}hree-\textbf{O}perator \textbf{S}plitting}
\newacronym{ADMM}{ADMM}{\textbf{A}lternating \textbf{D}irection \textbf{M}ethod of \textbf{M}ultipliers}
\newacronym{SLS}{SLS}{\textbf{S}ystem \textbf{L}evel \textbf{S}ynthesis}

\newacronym{CTM-s}{CTM-s}{
``Cell Transmission Model with service stations''}
\newacronym{ILC}{ILC}{Iterative Learning Control}
\newacronym{OB-ILC}{OB-ILC}{Optimization-based ILC}
\newacronym{MPC}{MPC}{Model Predictive Control}
\newacronym{ST}{ST}{Service Station}
\newacronym{TTT}{TTT}{Total Travel Time}
\newacronym{GT-MPC}{GT-MPC}{Ground-Truth MPC}
\IEEEoverridecommandlockouts                              

\usepackage{graphicx}
\usepackage{amsmath} 
\usepackage{amssymb}  
\usepackage{multirow}
\usepackage{cleveref} 
\creflabelformat{equation}{#2#1#3}
\crefname{equation}{}{}
\glsdisablehyper

\title{\LARGE \bf
Iterative Learning Control for Ramp Metering \\ on Service Station On-ramps
}

\author{Hongxi Xiang$^{1}$, Carlo Cenedese$^{2,3}$, Efe C. Balta$^{2,4}$, and John Lygeros$^{2}$
\thanks{
This work was supported as a part of NCCR Automation, a National Centre of Competence in Research, funded by the Swiss National Science Foundation (grant number 51NF40\_225155) \newline
{$^{1}$ Department of Mechanical and Process Engineering at ETH Z\"urich,  Z\"urich, Switzerland. Email: {\tt\footnotesize  hxiang@student.ethz.ch}.}
}%
\thanks{$^{2}$ Department of Information Technology and Electrical Engineering at ETH Z\"urich,  Z\"urich, Switzerland. Email: {\tt\footnotesize lygeros@control.ee.ethz.ch}.
}%
\thanks{$^{3}$  Delft Center for Systems and Control, Delft University of Technology, Delft, The Netherlands. Email: {\tt\footnotesize ccenedese@tudelft.nl}
}%
\thanks{$^{4}$ inspire AG, Z\"urich, Switzerland. Email: {\tt\footnotesize efe.balta@inspire.ch}}%
}

\begin{document}

\maketitle
\thispagestyle{empty}
\pagestyle{empty}

\begin{abstract}
Congestion on highways has become a significant social problem due to the increasing number of vehicles, leading to considerable waste of time and pollution. 
Regulating the outflow from the \acrlong{ST} can help alleviate this congestion. Notably, traffic flows follow recurring patterns over days and weeks, allowing for the application of \acrfull{ILC}. Building on these insights, we propose an \acrshort{ILC} approach based on the Cell Transmission Model with service stations (CTM-s). It is shown that \acrshort{ILC} can effectively compensate for potential inaccuracies in model parameter estimates by leveraging historical data.

\end{abstract}

\section{INTRODUCTION}

Traffic congestion in modern cities presents a major challenge for reducing emissions and combating climate change. Despite the emergence of alternative transportation options, roads remain the primary mode of transport for both people and goods in the US and EU~\cite{eutransportfigures}, 
with demand predicted to increase by 42\% by 2050~\cite{transport-in-eu}. Beyond the environmental impact and the related health risks, congestion leads to significant economic losses due to inefficiencies, such as increased fuel consumption and delayed delivery of goods \cite{ferrara2018freeway}.
Since naive solutions such as increasing capacity are not scalable, research has focused on optimizing the existing infrastructure. This approach requires a smaller investment and is adaptable to changing traffic conditions. 

The optimal operation of highway networks has long been a focus of research due to its significant impact on overall traffic. Numerous strategies have been proposed to reduce congestion \cite{papageorgiou2003review,papamichail2007motorway}, with on-ramp metering management emerging as one of the most effective ones \cite{papageorgiou2002freeway, vrbanic2021variable}.

While highway on-ramp management has been well-studied, the role of \glspl{ST} and their influence on traffic has gained attention only recently. In \cite{cenedese2022novel,cenedese2023optimal}, the authors proposed the \gls{CTM-s}, showing that (if located strategically), a \gls{ST} can reduce the peak traffic congestion on the highway stretch, acting as a damper preventing the congestion wave propagation. In~\cite{kamalifar:2024:metanet_ST} a second-order model based on METANET was proposed together with a simple control strategy based on ALINEA to regulate the flow of vehicles merging back into the mainstream from the \gls{ST}. These previous works did not consider the repetitive nature of the traffic demand, 
which makes the problem inherently well-suited for learning-based control strategies such as \gls{ILC}.

\gls{ILC} is a control strategy for iterative tasks, where the measurements of previous iterations are used for improving the control performance of subsequent iterations.
\gls{ILC} was originally formulated as an unconstrained tracking problem \cite{uchiyama1978formation,arimoto1984bettering}. To tackle constraints systematically, the \gls{OB-ILC} was then developed \cite{mishra2010optimization,owens2005iterative}, with theoretical results including convergence analyses shown in \cite{amann1996iterative}, \cite{schollig2009optimization}. These results do not consider noise or model mismatch, both crucial features of real-world applications. Adaptive ILC methods are developed to deal with parametric uncertainties in discrete-time repetitive systems \cite{yu2016data,oh2015stochastic}. \cite{meng2016robust} develops a scheme for discrete-time non-repetitive systems subject to iteration-varying unknown parameters and variable initial conditions. In \cite{nemec2017enhancing}, adaptation is achieved by combining \gls{ILC} with Reinforcement Learning. However, none of these methods provides robust constraint satisfaction guarantees. Recent work in \cite{liao2022robustness} develops an \gls{OB-ILC} scheme that uses a robust operator-theoretic framework to provide constraint satisfaction in the presence of noise and modeling errors for linear systems, also studied for iteration-varying systems~\cite{balta2024iterative}.

The main contributions of this paper are: (\textbf{i}) 
The first \gls{MPC} based ramp metering scheme using \gls{CTM-s} to control the flow exiting a \gls{ST} to reduce the mainstream congestion; (\textbf{ii}) a novel \gls{ILC} scheme for highway traffic control, where we systematically integrate previous traffic data into the controller design to improve performance. Our approach relaxes the knowledge and modeling requirements by providing means to compensate for parameter estimation errors. 
The efficacy of our \gls{ILC} scheme is validated via numerical studies demonstrating convergence to the correct parameter values and performance improvement. 

\begin{figure}[tb]
    \centering
    \includegraphics[width=\columnwidth]{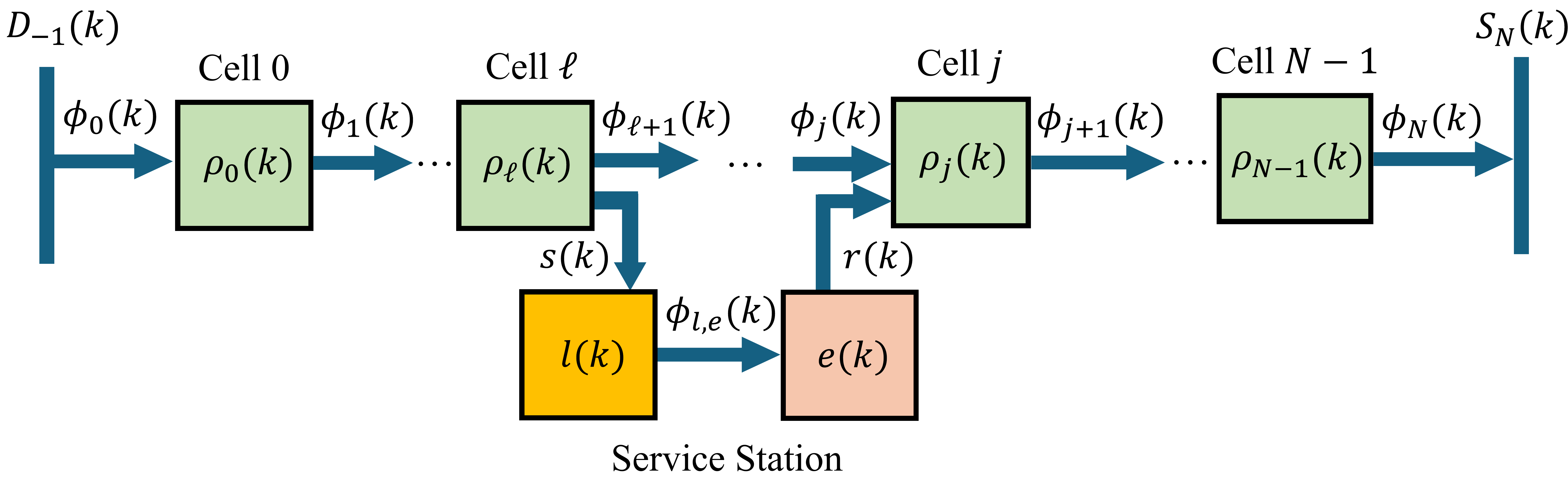}
    \caption{Illustration of \gls{CTM-s} model with one \gls{ST}.}
    \label{fig:ctms}
\end{figure}

\section{\gls{CTM-s} dynamics}
In this section, we adapt the \gls{CTM-s} model introduced in \cite{cenedese2022novel,cenedese2023optimal} to our context. In the model, time is discretized into intervals $[kT, (k+1)T)$ of length $T$ indexed by $k\in \mathbb{N}$, and the highway stretch is divided into $N$ consecutive cells indexed by $i \in \mathcal{N} = \{0, ..., N-1\}$, see~\cref{fig:ctms}. The vehicles exiting cell $i$ during the interval $k$ can either enter the subsequent cell $i+1$ or exit using an off-ramp if available. Similarly, new vehicles can enter cell $i$ from either cell $i-1$ or an on-ramp. The \gls{CTM-s} exploits the presence of on and off ramps to model the presence of a \gls{ST}. For simplicity, we assume there is only a single \gls{ST} identified by the pair $(\ell,j)$, where $\ell,j\in\mc N$, with $\ell$ and $j$ denoting the exit and merge cells to the \gls{ST}, respectively.

Differently from \cite{cenedese2022novel}, we model separately the vehicles actively using the service station $l(k)$ and those waiting in a queue while merging back from the \gls{ST}, $e(k)$. Moreover, we consider a highway stretch with no other ramps. It is important to emphasize that the following methodology can also be applied to more general cases, at the expense of more convoluted notation. Please refer to \cite{cenedese2022novel} for an in-depth discussion of the model and possible generalizations that relax these assumptions.

Next, we report the dynamics of the model using the variables in~\cref{tab:ctms}. We divide the dynamics into linear and non-linear, which will be useful for controller design.

\subsubsection{Linear dynamics}
The following dynamics guarantee the conservation of the number of vehicles and represent the cornerstone of the model. The density in cell $i$ evolves as
\begin{equation}
    \rho_i(k+1) = \rho_i(k)+\tfrac{T}{L_i}\left(\Phi_i^+(k)-\Phi_i^-(k)\right),
    \label{eq:ctms_dyn_rho}
\end{equation}
    where the total inflow and outflow are 
\begin{subequations}
	\begin{align}
		\Phi_i^+(k)& = \phi_i(k)+r_i(k),
		\label{eq:ctms_dyn_Phi+}\\
		\Phi_i^-(k)& = \phi_{i+1}(k)+s_i(k).
		\label{eq:ctms_dyn_Phi-}
	\end{align}
        \label{eq:flows}
\end{subequations}
Notice that $s_i(k)=s(k)= \beta \Phi_i^-(k-1)$ if $i=\ell$ and $0$ otherwise, similarly $r_i(k)=r(k)$ only if $i=j$ and $0$ otherwise. Vehicles spend on average $\delta$ time intervals ($\delta\; T$) receiving service within the station before joining the queue to exit, thus 
\begin{subequations}
\begin{align}
    l(k+1)& = l(k)+T(s(k) - \phi_{l, e}(k)),
    \label{eq:ctms_dyn_l}\\
    e(k+1)& = e(k)+T(\phi_{l,e}(k)-r(k)),
    \label{eq:ctms_dyn_e} \\
    \phi_{l, e}(k)& = s(k-\delta). \label{eq:ctms_hist}
\end{align}
\label{eq:ctms_ST_dynamics}
\end{subequations}

\subsubsection{Non-linear dynamics} 
The nonlinearities in the model are due to the flows among cells defined by the minimum between the upstream demand and downstream supply. The demand and supply for the cells and $\gls{ST}$ are defined as
\begin{subequations}
\begin{align}
    D_i(k) & = \min((1-\beta_i) v_i\rho_i(k), q_i^\text{max}), \label{eq:D_i} \\
    S_i(k) & = \min(w_i(\rho_i^\text{max} - \rho_i(k)), q_i^\text{max}),
    \label{eq:S_i}\\
    D^s(k) & = \min\left(\phi_{l, e}(k)+\tfrac{e(k)}{T}, r^\text{max}\right), \label{eq:D_s}
\end{align}
\label{eq:demand supply}
\end{subequations}
where in \eqref{eq:D_i}, $\beta_i = \beta$ if $i = \ell$ and $\beta_{i} = 0$ otherwise. 

For all $i\neq j$, the flow between cell $i-1$ and $i$ reads as
\begin{equation}
    \phi_i(k)= \min(D_{i-1}(k), S_i(k)). 
    \label{eq:simple_merge}
\end{equation}

On the contrary, if $i=j$, the supply of cell $i$ is shared by \gls{ST} and cell $i-1$, so
\begin{equation}
\phi_i(k) = \min (D_{i-1}(k), S^{\text{ms}}(k)),  
    \label{eq:station_merge_phi}     
\end{equation}
where $S^{\text{ms}}(k) \coloneqq \max (S_i(k)-D^s(k)), p^{\text{ms}} S_i(k))$, this formulation replaces the use of the \textit{middle} operator, used in~\cite[Eq.~3.34]{ferrara2018freeway}.
Similarly, we can compute $r(k)$ as follows
\begin{align}
    r(k) &= \min (D^s(k), S^{\text{ST}}(k)), \label{eq:station_merge_r}
\end{align}
where $S^{\text{ST}}(k) \coloneqq \max (S_{j}(k) - \phi_{j}(k), (1-p^{\text{ms}})S_{j}(k))$ denotes the remaining supply of cell $j$ available to the vehicles that aim to merge back.

As boundary conditions, we impose that $D_{-1}(k)$ is the exogenous upstream demand and that  $S_N(k) = +\infty$, hence no downstream bottleneck influences the flow on the considered highway stretch. 

\renewcommand*{\arraystretch}{1.25}
\begin{table}[t]
\begin{center}
\begin{tabular}{p{1.5cm}|c c l}
\toprule
\textbf{Name} &\textbf{Symbol} &\textbf{Unit} &\textbf{Description} \\
\midrule
\multirow{5}{1.5cm}{Cell parameters} 
& $L_i$ &[km]  & Cell length \\
&$v_i$ &[km/h]  & Free-flow speed\\
&$w_i$ &[km/h]  & Congestion wave speed\\
&$q_i^\text{max}$ &[veh/h]  & Flow capacity\\
&$\rho_i^\text{max}$ &[veh/km]  & Jam density\\

\cmidrule(lr){1-4}
\multirow{6}{1.5cm}{\gls{ST} parameters} 
&$l^\text{max}$ &[veh]  & Service station capacity\\
&$e^\text{max}$ &[veh]  & Queue length capacity\\
&$r^\text{max}$ &[veh/h]  & On-ramp capacity\\
&$\delta \ T$ &[h] & Average time spent at \gls{ST}\\
&$\beta$ &[-] & Split ratio at cell $\ell$\\
&$p^\text{ms}$ &[-] & Priority of mainstream\\

\cmidrule(lr){1-4}
\multirow{6}{1.5cm}{Cell variables} 
& $\rho_i(k)$ &[veh/km]  & Density of cell $i$ \\
&$\Phi_i^+(k)$ &[veh/h]  & Total inflow into cell $i$\\
&$\Phi_i^-(k)$ &[veh/h]  & Total outflow from cell $i$\\
&$\phi_i(k)$ &[veh/h]  & Flow from cell $i-1$ to $i$\\
&$D_i(k)$&[veh/h] &Demand of cell $i$\\
&$S_i(k)$&[veh/h] &Supply of cell $i$\\

\cmidrule(lr){1-4}
\multirow{6}{1.25cm}{\gls{ST} variables}
& $l(k)$ &[veh]  & Number of vehicles in \gls{ST}\\
&$e(k)$ &[veh]  & Queue length\\
& $s(k)$ &[veh/h]& Inflow to \gls{ST}\\
& $\phi_{l,e}(k)$ &[veh/h] & Flow from $l(k)$ to $e(k)$ \\
&$D^s(k)$&[veh/h] &Demand of \gls{ST}\\
&$r(k)$ &[veh/h]  & On-ramp flow from \gls{ST}\\

\cmidrule(lr){1-4}
\multirow{2}{1.5cm}{Boundary conditions}
&$D_{-1}(k)$&[veh/h] & Initial upstream demand\\
&$S_N(k)$&[veh/h] &Downstream supply\\
\bottomrule
\end{tabular}
\end{center}
\caption{\gls{CTM-s} parameters and variables}
\label{tab:ctms}
\end{table}
\renewcommand*{\arraystretch}{1}

\section{\gls{ST} ramp metering}
In this section, we formulate two controllers for ramp metering on the on-ramp allowing vehicles to exit the \gls{ST}. First, we formulate an \gls{MPC} that uses the \gls{CTM-s} to optimally restrict the \gls{ST}'s outflow to reduce the overall traffic congestion. Then, we augment this with an \gls{ILC} scheme, where we leverage the repetitiveness of the traffic demand. 

\subsection{\gls{MPC} formulation}
Ramp metering schemes based on \gls{MPC} have proven effective in real-life experiments~\cite{bellman:2006:ramp_metering_case_study}. 
In the case of \gls{ST}, the \gls{MPC} controller directly influences $D^s(k)$ via the input $r_c(k)$ that controls the maximum number of vehicles that can exit the \gls{ST} during the interval $k$. 
Therefore, the \gls{ST} demand in \eqref{eq:D_s} when the control is active becomes
\begin{align}
	D^s(k)& = \min\left(\phi_{l, e}(k)+\tfrac{e(k)}{T}, r^\text{max}, r_c(k)\right).
	\label{eq:D_s controlled}
\end{align}

The standard \gls{MPC} formulation requires defining a cost function $J$ to be minimized during the current time interval $k_0$, over a prediction horizon of length $K$, subject to constraints associated with the model dynamics and inputs. We consider a linear \gls{MPC} for efficient computation and approximate the nonlinear \gls{CTM-s} dynamics to attain only linear constraints.

\subsubsection{Constraints}
The nonlinear part of the \gls{CTM-s} dynamics arises from \eqref{eq:demand supply}--\eqref{eq:D_s controlled}. Inspired by \cite{gomes2006optimal}, we linearize these equations by relaxation.
For all $i\neq j$, we relax   \eqref{eq:simple_merge} as
\begin{align}
	\phi_i(k) \leq \min (D_{i-1}(k), S_i(k)),  \label{eq:relax_simple merge}
\end{align}
where the two are equivalent if \eqref{eq:relax_simple merge} is active. 
Substituting \eqref{eq:D_i} and \eqref{eq:S_i} into \eqref{eq:relax_simple merge} leads to
\begin{subequations}
\begin{align}
    \phi_i(k) &\leq (1-\beta_{i-1})v_{i-1}\rho_{i-1}(k), \\
        \phi_i(k) &\leq q_{i-1}^\text{max}, \\
        \phi_i(k) &\leq w_i(\rho_i^\text{max} - \rho_i(k)), \\
        \phi_i(k) &\leq q_i^\text{max}.
\end{align}
    \label{ineq:simple merge}
\end{subequations}
Similarly, for $i=j$ we relax  \eqref{eq:station_merge_phi}, \eqref{eq:station_merge_r} and \eqref{eq:D_s controlled} as
\begin{subequations}\label{ineq:station merge}
\begin{align}
    &\phi_i(k) \leq(1-\beta_{i-1})v_i\rho_{i-1}(k),\label{ineq:phi1}\\
    &\phi_i(k) \leq q_{i-1}^\text{max},\label{ineq:phi2}\\
    &r(k) \leq \phi_{l, e}(k) + e(k)/T,\label{ineq:r1}\\
    &r(k) \leq r^\text{max}, \label{ineq:r2}\\
    &\phi_i(k) + r(k) \leq w_i(\rho_i^\text{max} - \rho_i(k)), \label{ineq:sum1}\\
    &\phi_i(k) + r(k) \leq q_i^\text{max},\label{ineq:sum2}
\end{align}
\end{subequations}
where \eqref{ineq:phi1} and \eqref{ineq:phi2} constrain $\phi_i(k)$ by the demand of cell $i-1$, while \eqref{ineq:r1} and \eqref{ineq:r2} constrain $r(k)$ by the demand of \gls{ST}. Finally, \eqref{ineq:sum1} and \eqref{ineq:sum2} indicate that the sum of the two is constrained by the supply of the next cell $i$. 

Notice the approximation above ignores $p^\text{ms}$, i.e., we only consider the first terms in both $S^{\text{ms}}(k)$ and $S^{\text{ST}}(k)$. In practice, $p^\text{ms}$ is usually close to $1$ hence the mainstream flow has a higher priority than the flow from the \gls{ST}. We promote this behavior in our controller by weighting $\phi_i(k)$ more than $r(k)$ in the cost function design, as detailed in the next section.

Among the controller's goals, we also include limiting the maximum queue length at the exit of the \gls{ST}. This can be imposed via the following constraint
\begin{align}
    e(k) \leq e^\text{max}, \label{ineq:emax}
\end{align}
where $e^\text{max}$ is the fixed maximum queue length.
\subsubsection{Cost function}
The primary control objective is to minimize the traffic congestion on the highway stretch.
As the congestion metric, we use the \gls{TTT} over the whole prediction horizon, which is a widely used metric in highway control in the literature, e.g., see \cite[Ch.8]{ferrara2018freeway}, and it is defined as
\begin{align}
    \mathrm{TTT}(k_0,k_0+K)\coloneqq {\textstyle \sum_{k=k_0}^{k_0 + K} \sum_{i=0}^{N-1} \rho_i(k) L_i.}
    \label{eq:TTT}
\end{align}

\begin{figure}[tb]
    \centering
    \includegraphics[width=\columnwidth]{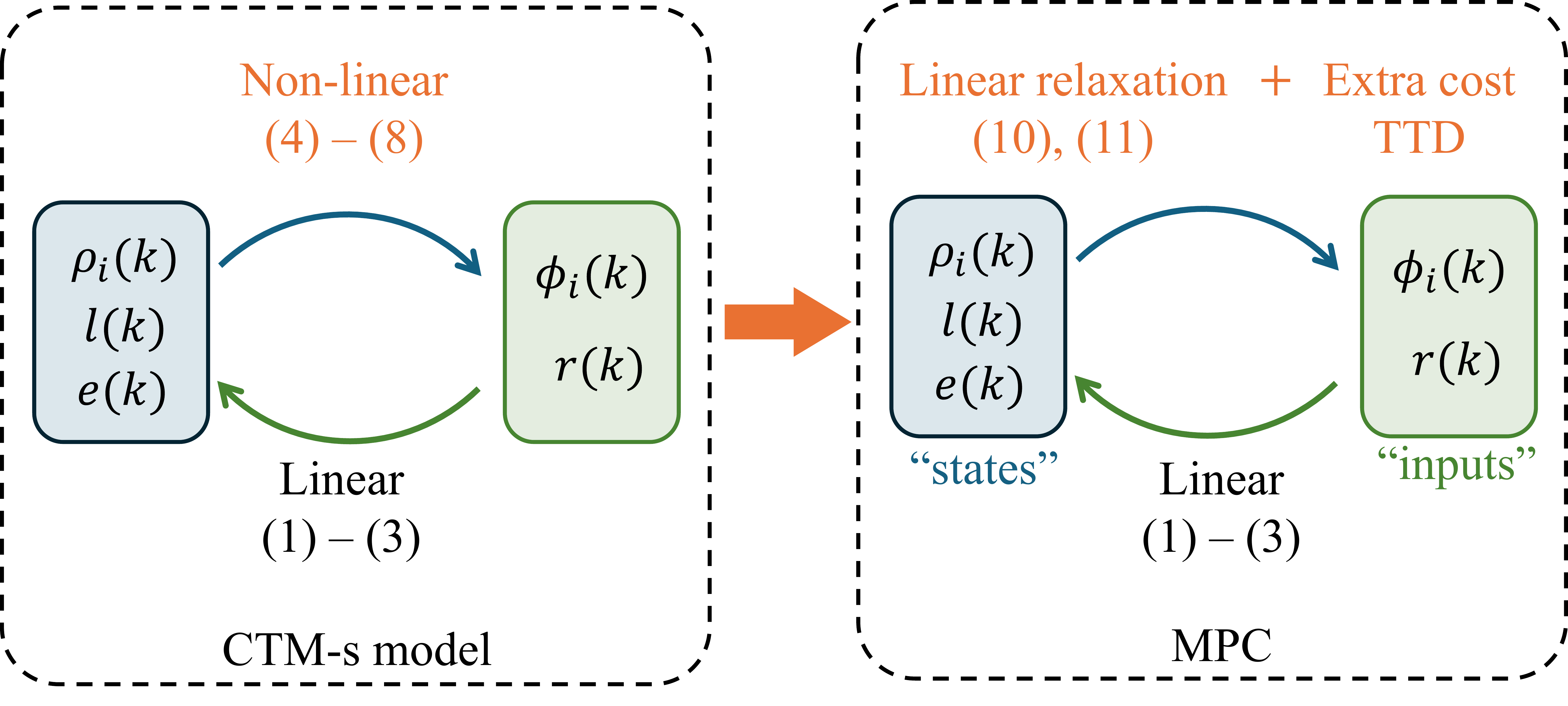}
    \caption{Relations between CTM-s model and MPC formulation. The non-linear dynamics \eqref{eq:demand supply} - \eqref{eq:D_s controlled} are relaxed as linear constraints \eqref{ineq:simple merge}, \eqref{ineq:station merge} with an extra cost term TTD. Moreover, the variables in blue boxes are known at time step $k$, while those in green boxes are only known after $k+1$. Thus, they are defined as states and inputs, respectively.}
    \label{fig:states and inputs}
\end{figure}

To compensate for the relaxed constraints, we design a cost function to push the variables against the constraint \eqref{ineq:simple merge} - \eqref{ineq:station merge}. This approach leads to optimal solutions that satisfy a subset of the constraints with equality, which, in turn, satisfy the original non-linear \gls{CTM-s} dynamics. To this end, we maximize the flows $\phi_i(k)$ and $r(k)$.
Therefore, the cost $J$ minimized by the \gls{MPC} becomes
\begin{align}
    \mathrm{TTT}(k_0,k_0+K)-\lambda  \sum_{k=k_0}^{k_0 + K - 1} \left(w_r r(k) + \sum_{i=0}^{N} \phi_i(k) L_{i-1} \right),
    \label{eq:linear cost}
\end{align} 
where $\lambda>0$ and small $w_r$ ($0<w_r<L_{j-1}$) gives priority to the mainstream flow $\phi_j(k)$ when merging. 
The last term in \eqref{eq:linear cost} is called Total Travel Distance (TTD), see \cite[Ch.8]{ferrara2018freeway}. 
Since there is no cell $-1$, we set $L_{-1}$ to a predefined weight. 
Note that during the control horizon $[k_0 T, (k_0+K)T]$, the flows $\phi_i(k_0+K)$ and $r(k_0+K)$ have not taken place yet, since by definition, they are flows during $[(k_0+K)T, (k_0+K+1)T)$. Therefore, the flows are only summed up to $K_0+K-1$ instead of $k_0+K$. 
For the same reason, in the following section, we define $\phi_i(k)$ as part of the inputs in the \gls{MPC} formulation, even though we only regulate $r(k)$, since there are no initial values for them to propagate the dynamics, see \cref{fig:states and inputs}. 

\subsubsection{Overall formulation} 
To formulate the linear \gls{MPC},
we first organize the \gls{CTM-s} variables as states $x(k)  \coloneqq [\rho_0(k), \hdots \rho_{N-1}(k), l(k), e(k)]^\top\in \mathbb{R}^{N+2}$, and inputs $u(k) \coloneqq [\phi_0(k), \hdots, \phi_N(k), r(k)]^\top\in \mathbb{R}^{N+2} $. 
Then the collection of the variables over the prediction horizon is denoted by $x_{k_0}  \coloneqq\mathrm{col}\left(\{x(k_0+k)\}_{k=0}^K\right)\in \mathbb{R}^{(N+2) \times (K+1)}$ and $u_{k_0} \coloneqq\mathrm{col}\left(\{u(k_0+k)\}_{k=0}^{K-1}\right)\in \mathbb{R}^{(N+2) \times K} $, ${\phi_{l, e, k_0}}\coloneqq\mathrm{col}\left(\{\phi_{l, e}(k_0+k)\}_{k=0}^{K-1}\right) \in \mathbb{R}^{K}$. 

The \gls{MPC} at time at time $k_0$ reads as
\begin{subequations}\label{program: MPC lifted}
    \begin{align}
        \min_{u_{k_0}}  J(u_{k_0}) & =\tfrac{a}{2} \lVert x_{k_0} - x_r \rVert ^2_Q + c_x^\top x_{k_0} - \ c_u ^\top u_{k_0},\label{MPC obj lifted} \\ 
        \text{s.t.} \quad  & x_{k_0} = x_{\text{init}, k_0} + G u_{k_0} + G_h \ {\phi_{l,e, k_0}}, \label{MPC eq cons lifted} \\ 
        & A_{x} x_{k_0} + A_u u_{k_0} \leq b_{k_0}, \label{MPC ineq cons lifted} \\
        & x_{k_0} \geq 0, u_{k_0} \geq 0. \label{MPC nonnegative lifted}
	\end{align}
\end{subequations}

In the objective function \eqref{MPC obj lifted}, the two linear cost terms correspond to \eqref{eq:linear cost}. Moreover, we augment the cost with an additional quadratic term to provide necessary gradient information for the \gls{ILC} scheme devised in the next section. However, this quadratic term should be designed carefully since it potentially diverges the optimal solution from satisfying the subset of relaxed constraints \eqref{ineq:simple merge} and \eqref{ineq:station merge} with equality. 
Since we are to minimize $x$, we set $x_r:=0$ in our implementation. We further define a diagonal $Q\succ 0$, where its diagonal elements associated with $\rho_i(k)$, $l(k)$, and $e(k)$ are set to $w_{\rho} L_i/ \rho_i ^ \text{max}$, $w_l / l ^ \text{max}$, and $w_e / e^\text{max}$, with weights $w_{\rho},\,w_l,\,w_e>0$ normalized by $\rho_i ^ \text{max}$, $l ^ \text{max}$, and $e^\text{max}$, respectively. 
This ensures that the quadratic terms behave approximately linearly to conserve the tightness of the optimal solution at the subset of relaxed constraints \eqref{ineq:simple merge} and \eqref{ineq:station merge} as much as possible. The predefined $a>0$ is then used to achieve a reasonable trade-off between the gradient information and inaccuracy induced by relaxation.

In  \eqref{MPC eq cons lifted}, $x_{\text{init}, k_0}\coloneqq\mathbb{1}\otimes x(k_0)$ with $\mathbb{1}$ as a vector of ones of the correct dimension. The constraints \eqref{eq:ctms_dyn_rho} -- \eqref{eq:ctms_ST_dynamics},  \eqref{ineq:simple merge} -- \eqref{ineq:emax}  are included in compact form in \eqref{MPC eq cons lifted}, \eqref{MPC ineq cons lifted} and \eqref{MPC nonnegative lifted}, where $G,G_h,A_x,A_u$ are matrices of the correct dimensions.

In practice, the time sampling interval $T$ is a design choice, and highway parameters $L_i$, $v_i$, $w_i$, $q_i^\text{max}$, $r^\text{max}$ are constant and can usually be measured precisely. However, the split ratio $\beta$, average waiting time $\delta$, and future initial demand $D_{-1}(k)$ depend on the drivers and are difficult to estimate. Therefore, \eqref{program: MPC lifted} is solved by using the best estimates $\beta_{\mathrm{es}}$, $\delta_{\mathrm{es}}$, and $D_{-1, \mathrm{es}}(k)$ of the corresponding parameters, i.e., \eqref{MPC eq cons lifted} and \eqref{MPC ineq cons lifted} are substituted by
\begin{subequations}\label{estimated_cons}
    \begin{align}
        & x_{k_0} = x_{\text{init}, k_0} + M u_{k_0} + G_h \ {\phi_{l,e, k_0, \mathrm{es}}}, \label{MPC eq cons lifted est} \\ 
        & A_{x, \mathrm{es}} x_{k_0} + A_u u_{k_0} \leq b_{\mathrm{es}}, \label{MPC ineq cons lifted est}
	\end{align}
\end{subequations}
where $M$, $A_{x, \mathrm{es}}$, $b_{\mathrm{es}}$, and ${\phi_{l,e, k_0, \mathrm{es}}}$ are estimates of $G$, $A_x$, $b$, and ${\phi_{l,e, k_0}}$, respectively, based on estimates $\beta_{\mathrm{es}}$, $\delta_{\mathrm{es}}$, and $D_{-1, \mathrm{es}}(k)$. 

After solving the MPC problem, we set the obtained optimal input $r^*(k)$ as the control input $r_c(k)$ in \eqref{eq:D_s controlled}. Typically, in an MPC setting, the control input is updated at every time step. However, empirical experiments suggest that increasing the update interval has minimal impact on control performance. In our implementation, we update the control input every $p$ steps to reduce the computation load.

\subsection{\gls{ILC} formulation}
\subsubsection{Motivation for \gls{ILC}}
As we will see in simulations, poor estimates $\beta_{\mathrm{es}}$, $\delta_{\mathrm{es}}$, and $D_{-1, \mathrm{es}}(k)$ can significantly undermine performance. 
To improve performance over time, we exploit the approximately repetitive nature of the problem. Specifically, the traffic pattern on the highway repeats from day to day and from week to week, enabling us to ``learn'' and ``improve'' the estimates of the unknown parameters based on previous data. 

\subsubsection{The control framework}
\begin{figure}[tb] 
    \centering
\includegraphics[width=\columnwidth]{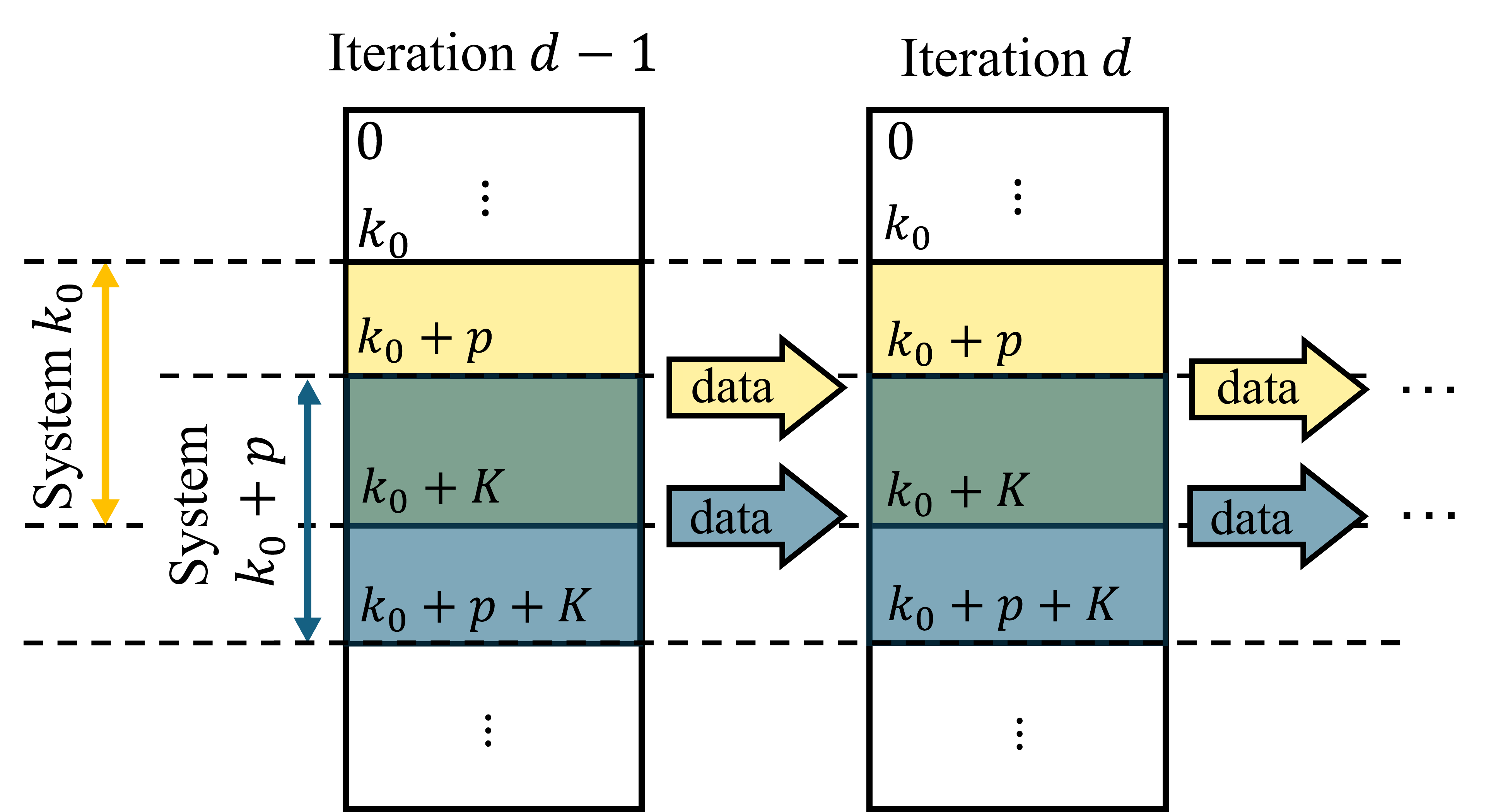}
    \caption{General framework of applying \gls{ILC} to highway traffic. Yellow and blue boxes denote \textit{system $k_0$} and \textit{system $k_0+p$} (with horizon length $K$), respectively. We compute the control inputs for system $k_0$ on iteration $d$ by \eqref{modified ILC} using data of the same system on iteration $d-1$. Then we apply the control inputs until updated at time step $k_0+p$ for the next system.}  
    \label{fig:ILC}
\end{figure}

\Cref{fig:ILC} illustrates how the \gls{ILC} framework is implemented for highway traffic control. 
For simplicity, we assume that the traffic pattern repeats daily, though our approach can be easily adapted to any period. 

We implement \gls{ILC} in a receding-horizon fashion. 
For computing control actions at time step $k_0$ of iteration $d$, i.e., the $d$-th day, we use the data of time steps $k \in [k_0, k_0+K]$ collected during the previous iteration $d-1$. 
We update control inputs every $p$ time step. Therefore, at time step $k_0+p$, the control inputs are recomputed using the data of $k \in [k_0+p, k_0+p+K]$ of the previous iteration. This process repeats until the entire desired control period is covered.

We call the traffic between $k_0$ and $k_0+K$ (denoted by the yellow boxes in~\cref{fig:ILC}) \textit{system $k_0$}, which corresponds to the nominal problem \eqref{program: MPC lifted} at $k_0$. 
Similarly, we call the traffic between $k_0+p$ and $k_0+p+K$ (denoted by blue boxes in~\cref{fig:ILC}) \textit{system $k_0 + p$}, and so on.

Note that we iterate horizontally rather than vertically. This is because the repetitive pattern of highway traffic occurs simultaneously on each iteration (e.g., day), while the traffic during a single iteration (e.g., during one day) can vary largely at different time steps. For example, the traffic demand during the morning peak strongly differs from one in the evening, see~\cref{fig:initial demand}, on the contrary, the morning peak traffic demand on one day is usually similar to one of the previous day.

We propose an \gls{ILC} formulation based on~\cite{liao2022robustness} and focus on a particular system $k_0$ and omit the subscript ``$k_0$'' if there is no ambiguity.
The subscript $d$ denotes the iteration index.

\subsubsection{Formulation}
Based on \eqref{program: MPC lifted}, we propose the following \gls{ILC} formulation.
\begin{subequations}\label{modified ILC}
    \begin{align}
    \mathop{{\text{min}}}_{v}  \ \tfrac{1}{2} &\|v-u_{d-1}\|^2_W + \alpha\; v ^ {\top} {F}(x_{d-1}, u_{d-1}), \label{eq:ILC obj}\\
         s.t. \quad  {x}_d = & M v + x_{\text{init}, d}\! +\! {x}_{d-1}\! -\! M{u}_{d-1}\! -\! x_{\text{init},d-1},  \label{ILC eq cons modified}\\ 
        & A_{x,es} {x}_d + A_u v \leq b_{d-1}, \label{ILC ineq cons modified}  \\
        & {x}_d \geq 0 \quad v \geq 0,  \label{ILC nonnegative modified}
    \end{align}
\end{subequations}
where $u_{d-1}$, $x_{d-1}$ are measurements from the previous iteration.  
Additionally $b_{d-1}$ uses previous data ${D_{-1, d-1}(k)}$ and ${\phi_{l,e, d-1}}(k)$. 
For the objective \eqref{eq:ILC obj}, $W := M^\top Q M \succ 0$ is a preconditioner matrix. Note that in our formulation, $M$ has full column rank, therefore it is guaranteed that $W \succ 0$ as long as $Q \succ 0$. 
$\alpha>0$ is a predefined weight and $F$ is an approximation of the gradient of the nominal objective function \eqref{MPC obj lifted}, defined next. 

For iteration $d$, by replacing \eqref{MPC obj lifted} into \eqref{MPC eq cons lifted}, we attain
\begin{align}
    J_d(u) = \frac{1}{2} u^\top H u + f_d^\top u + c,
\end{align}
where $H := a \; G ^\top Q G$, $f_d:=a \; G^\top Q (w_d - x_r) + G^\top c_x - c_u$, $w_d:=G_h{\phi_{l, e}}_d + x_{\text{init,d}}$, and $c:=G^\top c_x - c_u$. Then we have
\begin{align}
    &\nabla J_d(u) = H u + f_d  \nonumber\\
    &= a\; [G^\top Q(Gu+w_d-x_r)] + G^\top c_x - c_u.
    \label{eq:gradient}
\end{align}
Since we do not know $G$, and $w_d$ exactly, we define
\begin{align}
    F(x_{d-1}, u_{d-1})\! :=\! a\;[M^\top Q(x_{d-1}-x_r)]\! +\! M^\top c_x\! -\! c_u
    \label{eq:gradient_approx}
\end{align}
as an approximation of the gradient of the objective function of iteration $d$ evaluated at input $u$ of iteration $d-1$, as $$\nabla J_d(u_{d-1}) = a \;[G^\top Q(Gu_{d-1}+w_{d}-x_r)] + G^\top c_x - c_u,$$
where we replaced the unknown $G$ with its estimate $M$ and $G u_{d-1} + w_d$ by $x_{d-1}$. This is equivalent to replacing $w_d$ by $w_{d-1}$, since $x_{d-1} = G u_{d-1} + w_{d-1}$, according to \eqref{MPC eq cons lifted}. 

Next, we provide the intuition behind constraint \eqref{ILC eq cons modified}. Let $x_{d, GT}$ denote the ground truth values computed by using exact parameters, i.e., $x_{d, GT} = x_{\text{init, d}} + G v + G_h {\phi_{l,e, d}}$, then the computation error when using \eqref{ILC eq cons modified} reads as
\begin{align}
    \text{err}_d :=& x_{d, GT} - x_d \notag \\
    =& x_{\text{init, d}} + G v + G_h {\phi_{l,e,d}} \nonumber \\
    & -  (M v + x_{\text{init}, d} + {x}_{d-1} - M{u}_{d-1} - x_{\text{init},d-1}) \nonumber\\
    =& (G-M)(v-u_{d-1}) + G_h ({\phi_{l,e, d}} - {\phi_{l,e, d-1}}),
    \label{eq:intuition}
\end{align}
where the last equality comes from $x_{d-1} = Gu_{d-1} + x_{\text{init}, d-1} + G_h {\phi_{l,e, d-1}}$. Therefore, since from \eqref{eq:ILC obj}  we try to minimize the distance between $v$ and  $u_{d-1}$, then the estimated error $G-M$ will be compensated. 

After computing \eqref{modified ILC}, we apply the optimal solution $r^*(k)$ for $p$ time intervals as $r_c(k)$, and then compute a new control input sequence by solving a new instance of \eqref{modified ILC} associated with the next system.

\section{Simulations}
\subsection{Simulations setup}
We consider a highway stretch identified by the parameters in~\cref{tab:simparams}. 
There is a notable reduction of $w_i$ and $q_i^{\text{max}}$ from cell $8$ to $9$, which creates a bottleneck. The \gls{ST} is placed upstream of the bottleneck, between cells $\ell=4$ and $j=6$, to affect traffic congestion induced by the bottleneck. For further details on the optimal \gls{ST} placement refer to \cite{cenedese2023optimal}.

\begin{figure}[tb]
    \centering
    \includegraphics[width=\columnwidth]{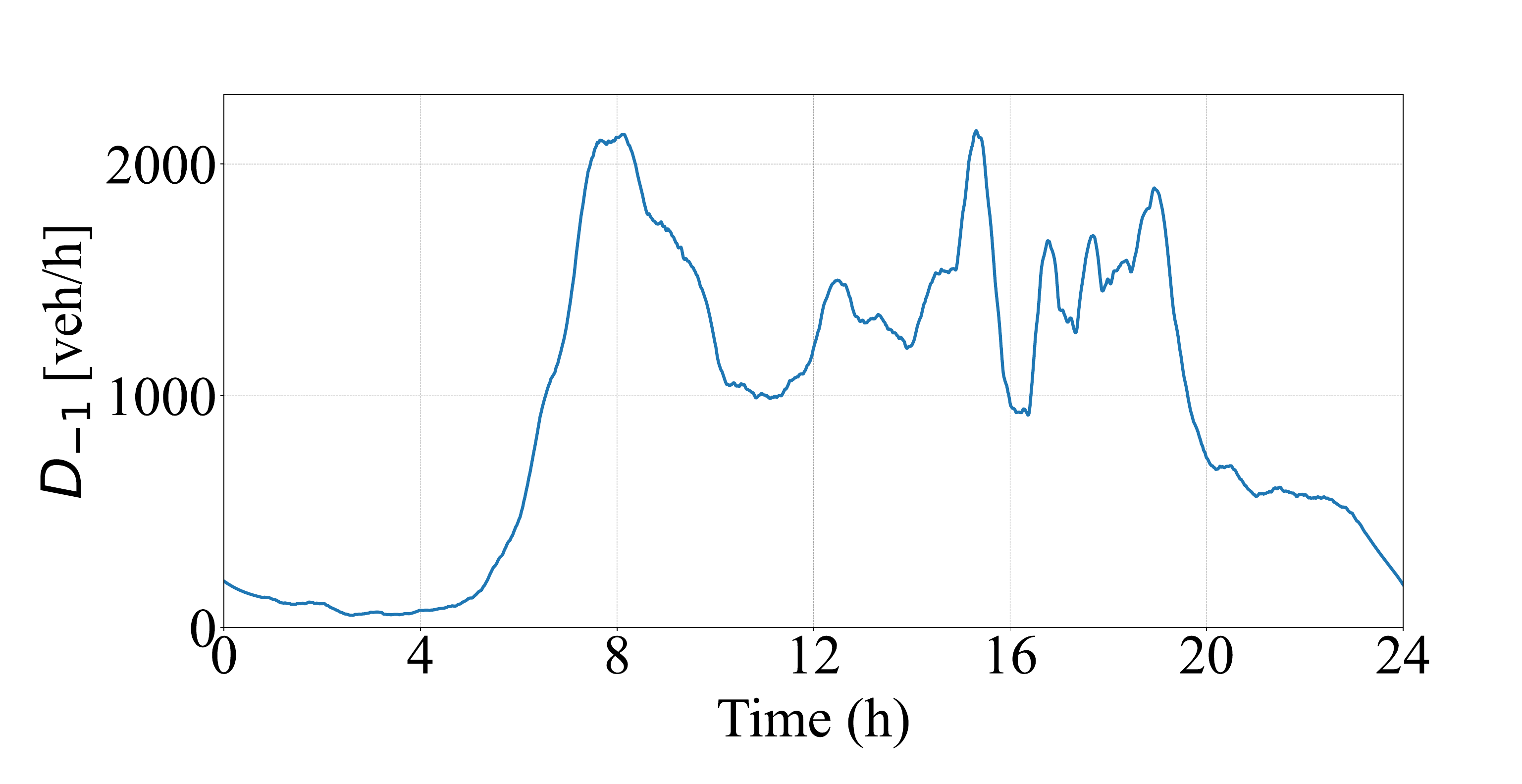}
    \caption{Initial upstream demand $D_{-1}(k)$.
    }
    \label{fig:initial demand}
\end{figure}

The upstream demand $D_{-1}(k)$ is as in~\cref{fig:initial demand} that is a scaled version of the flow along the A2 highway in the Netherlands during a standard weekday, extracted from the \textit{Nationaal Dataportaal Wegverkeer} \cite{ndw_traffic_data}. \eblue
In the simulations, $D_{-1}(k)$ is set to be the same for all days, one iteration corresponds to one day.
We focus only on the morning peak from $07:00$ to $10:00$, the initial and final time of the considered peak are denoted by $t_{\rm s}$ and $t_{\rm e}$, respectively. 
We model the error in the initial estimates of $\beta$, $\delta$, and $D_{-1}(k)$, by the scaling factors $r_\beta,\,r_\delta,\,r_D\geq0$. 
The estimated parameters are 
\begin{align}
\beta_{\mathrm{es}}=r_\beta\beta,\quad \delta_{\mathrm{es}}=r_\delta\delta,\quad  D_{-1, \mathrm{es}}(k)=r_D D_{-1}(k), \label{eq:estimates}
\end{align}
where a scaling greater or smaller than $1$ denotes an over- or underestimation of the corresponding parameter.
Our goal is to show that over the iterations, the \gls{ILC} compensates for this estimation error and achieves performance similar to the ones of the \gls{GT-MPC}, that is
the \gls{MPC}  in \eqref{program: MPC lifted} with the correct parameters in \cref{tab:simparams} and $D_{-1}(k)$.

\subsection{Controllers and performance metrics} 
We consider six different scenarios where the parameters in \eqref{eq:estimates} take values as in~\cref{tab:r_val}.
\begin{table}[tb]
\begin{center}
\begin{tabular}{c | c c c }
\toprule
&$\mathbf{r_{\beta}}$&$\mathbf{r_{\delta}}$&$\mathbf{r_{D}}$\\
\midrule
Underest.
&0.8&0.8&0.8\\
Overesti.
&1.2&1.2&1.2\\
\bottomrule
\end{tabular}
\end{center}
\caption{The scaling factors used in the six simulations scenarios to impose under or overestimation errors.}
\label{tab:r_val}
\end{table}
Notice that, during the first day ($d=0$), the \gls{ILC} scheme cannot be applied, as there is no prior day's data available. Therefore, an \gls{MPC} is used with parameters $\beta_{\mathrm{es}}$, $ \delta_{\mathrm{es}}$, and $ D_{-1,{\mathrm{es}}}$. 

Since our primary goal is to reduce congestion, we use TTT$(t_{\rm s},t_{\rm e})$ as  main evaluation metric. 
The controller trad offs between alleviating congestion and maintaining a low waiting time at the \gls{ST} exit.
Thus, we consider two additional metrics the Total Waiting Time (TWT) and the Total Time Spent (TTS), defined as
\begin{subequations}
\begin{align}
    \mathrm{TWT}(t_{\rm s},t_{\rm e}) & \coloneqq \sum_{k=t_{\rm s}}^{t_{\rm e}} e(k),  \label{eq:metricsTWT} \\
    \mathrm{TTS}(t_{\rm s},t_{\rm e}) &\coloneqq \mathrm{TTT}(t_{\rm s},t_{\rm e}) + \mathrm{TWT}(t_{\rm s},t_{\rm e}). \label{eq:metricsTTS} 
\end{align}
    \label{eq:metrics}
\end{subequations}
TWT measures the waiting time for vehicles at the \gls{ST} exit, while TTS represents the overall time for traveling and waiting. Moreover, to better assess the convergence performance of \gls{ILC} to \gls{GT-MPC}, we define the following relative indices \eblue
\begin{subequations}
\label{eq:delta_metrics}
    \begin{align}
    \Delta_\mathrm{TTT} & \coloneqq \mathrm{TTT}^{\mathrm{ILC}}(t_{\rm s},t_{\rm e})- \mathrm{TTT}^{\mathrm{GT-MPC}}(t_{\rm s},t_{\rm e}),\label{eq:DeltaTTT} \\
    \Delta_\mathrm{TWT} & \coloneqq \mathrm{TWT}^{\mathrm{ILC}}(t_{\rm s},t_{\rm e})- \mathrm{TWT}^{\mathrm{GT-MPC}}(t_{\rm s},t_{\rm e}),\label{eq:DeltaTWT} \\
    \Delta_\mathrm{TTS} & \coloneqq \mathrm{TTS}^{\mathrm{ILC}}(t_{\rm s},t_{\rm e})- \mathrm{TTS}^{\mathrm{GT-MPC}}(t_{\rm s},t_{\rm e}),\label{eq:DeltaTTS} 
	\end{align} 
\end{subequations}
where the superscript indicates the controller used.
Finally, we define the maximum rate of violating the queue length constraint \eqref{ineq:emax} over  $[t_s, t_e]$ as
\begin{align}
    \Delta_{e^{\text{max}}}(t_{\rm s}, t_{\rm e}) &\coloneqq \max_{k \in [t_{\rm s}, t_{\rm e}]} \left( \frac{\max\left(e(k) - e^{\text{max}}, 0\right)}{e^{\text{max}}} \right).
    \label{eq:metricsvio}
\end{align}
If there is never a violation, then  $\Delta_{e^{\text{max}}}=0$.

\renewcommand*{\arraystretch}{1.25}
\begin{table}[tb]
\begin{center}
\begin{tabular}{c | c c c c c c}
\toprule
\multirow{16}{*}{\text{\textbf{Cells}}}
& ${i}$ &${L_i}$ &${v_i}$ &${w_i}$ &${q^{\text{max}}_i}$ &${\rho_{\text{max}}^i}$\\
\cline{2-7}
&0&0.65&103&31&1870&79\\
&1&0.56&103&25&1735&86\\
&2&0.61&103&33&1876&75\\
&3&0.23&103&26&1757&84\\
&4&0.34&103&33&1780&71\\
&5&0.54&103&35&1847&71\\
&6&0.29&103&38&1985&72\\
&7&0.31&103&40&2092&73\\
&8&0.59&103&40&2002&69\\
&9&0.6&96&29&1714&77\\
&10&0.41&96&29&1705&76\\
&11&0.2&103&33&1845&74\\
&12&0.7&103&35&1924&74\\
&13&0.53&104&30&1774&77\\
&14&0.51&103&27&1789&83\\
\midrule
\multirow{2}{*}{\text{\textbf{\gls{ST}}}}
&$l^\text{max}$&$e^\text{max}$&$r^\text{max}$&$\delta$&$\beta$&$p^\text{ms}$\\
\cline{2-7}
&400&20&1500&480&0.1&0.9\\
\midrule
\multirow{4}{*}{\text{\textbf{Other}}}
&T [s]&$K$&$p$&$\lambda$&$a$&$\alpha$\\
\cline{2-7}
&$10$&$90$&$30$&$0.5$&$1$&$1$ \\
\cline{2-7}
&$w_\rho$&$w_e$&$w_l$&$w_r$&$L_{-1}$&\\
\cline{2-7}
&$1$&$0.1$&$0.05$&0.1&0.5& \\
\bottomrule
\end{tabular}
\end{center}
\caption{Parameters of the \gls{CTM-s} and the controller used in the simulations.}
\label{tab:simparams}
\end{table}
\renewcommand*{\arraystretch}{1}

\subsection{Results}
\begin{table}[tb]
\begin{center}
\begin{tabular}{ c | c c c c }
\toprule
&\textbf{TTT}&\textbf{TWT}&\textbf{TTS}&$\mathbf{\Delta_{e^{\text{max}}}}$\\
\midrule
Uncontrolled
&358.49&0.55&359.04&0\\
GT-MPC
&344.64&14.63&359.27&0\\
\bottomrule
\end{tabular}
\end{center}
\caption{Performance of uncontrolled stretch and the one under GT-MPC.}
\label{tab:GTMPC_uncontrolled}
\end{table}

\begin{figure}[tb]
    \centering
    \includegraphics[width=\columnwidth]{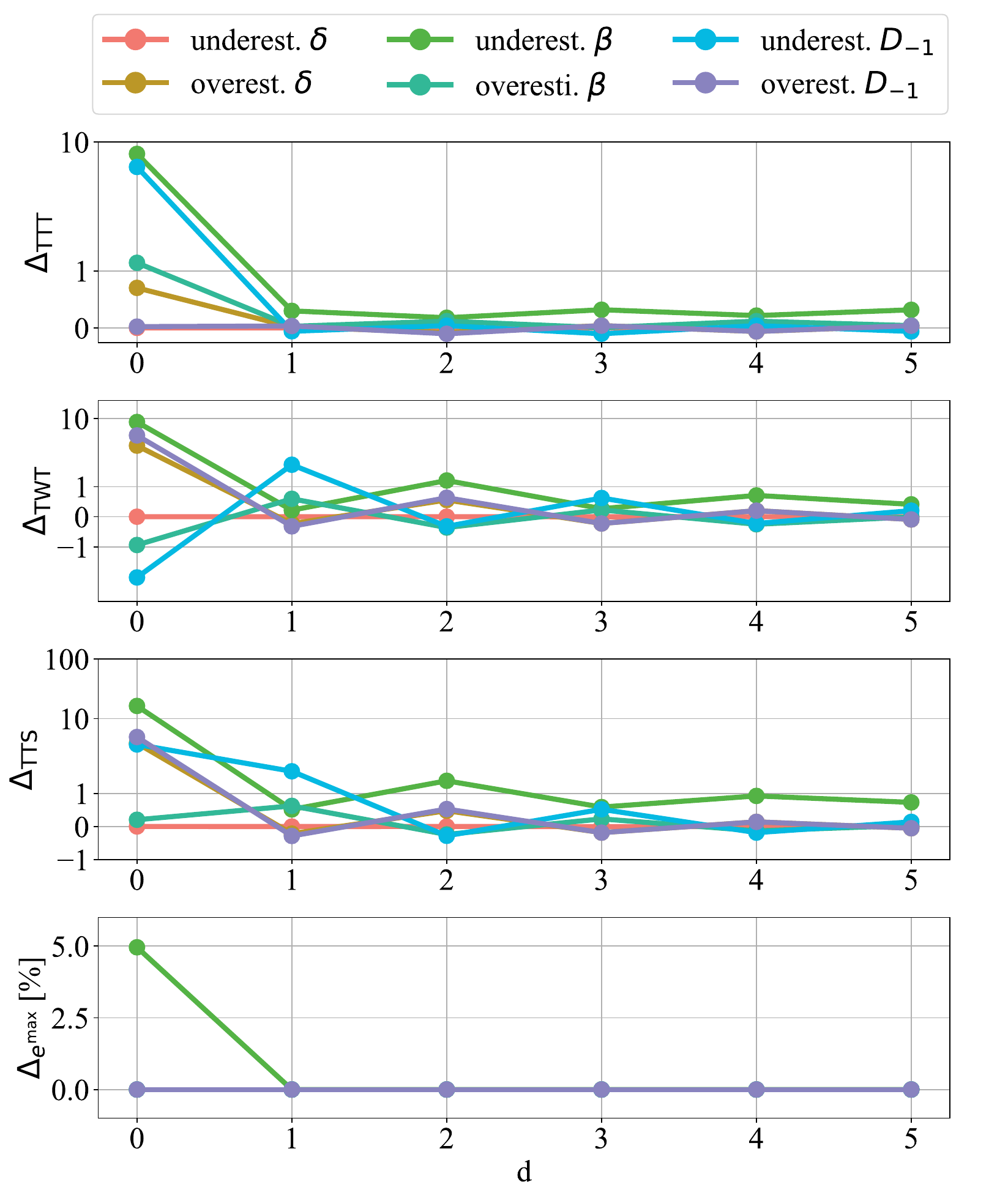}
    \caption{Relative performance of the \gls{ILC} scheme over $5$ iterations with the scenarios in Table~\ref{tab:r_val}. 
    }
    \label{fig:metrics}
\end{figure}

\cref{tab:GTMPC_uncontrolled} shows the results for the highway without control and under the control of \gls{GT-MPC}. Notice that TTS are almost equal in both cases, indicating the trade-off between TTT and TTS, as expected. Thanks to its precise knowledge of the parameters, \gls{GT-MPC} is effective, as it reduces congestion (attain lower TTT) while maintaining $\Delta_{e^{\text{max}}} = 0$.  

\Cref{fig:metrics} shows the relative performance of \gls{ILC} in the six scenarios summarized in~\cref{tab:r_val}. 
Note that the first three sub-figures are presented on a logarithmic scale.
At iteration $0$, since \gls{MPC} with estimated parameters is applied, the positive values of $\Delta_\mathrm{TTT}$ suggest that the control effect of reducing congestion can be undermined by imprecise knowledge of the parameters. In some cases (e.g., underestimating $\beta$), due to wrong estimates, the controller can result in both positive $\Delta_\mathrm{TTT}$ and $\Delta_\mathrm{TWT}$ simultaneously, 
leading to greater overall TTS. Moreover, the queue length constraints can be violated as well. 
Remarkably, by exploiting previous data at subsequent steps,  \gls{ILC} is able to converge rapidly towards \gls{GT-MPC} (within one or two iterations) in all cases, achieving lower $\Delta_\mathrm{TTT}$ (therefore lower TTT), as indicated by the fast convergence of all curves to a neighborhood of $0$ value. Moreover, the queue length constraint can always be satisfied. This shows the effectiveness of our \gls{ILC} approach.

\section{CONCLUSIONS}
\gls{ILC} tailored to the \gls{CTM-s} model is an effective algorithm to perform highway traffic control and mitigate the presence of parameters' identification errors. Numerical experiments show that by leveraging the repetitive traffic pattern, our \gls{ILC} approach converges to the optimal \gls{MPC} controller with ground-truth parameters.
Our \gls{ILC} method recovers from the incorrect initial estimations in a few iterations and outperforms \gls{MPC} under the same conditions, remarkably, even when the parameter values differ by $20\%$ from the correct ones. 

This study focuses on a single highway with one service station. An interesting extension is to use this approach to control more complex road networks with multiple stations, where coordinated control may yield better performance. 
Another potential direction is developing a nonlinear \gls{ILC} to address model mismatch introduced by the linear relaxation of the \gls{CTM-s} model, which inherently contains nonlinear dynamics. 
Our current formulation requires an extra cost term TTD, which can potentially diverge the controller from the original goal of reducing TTT. With non-linear formulation, we might skip relaxation and remove TTD, and therefore reduce congestion more effectively.

\addtolength{\textheight}{-12cm}




\bibliography{bibliography}
\bibliographystyle{ieeetr}

\end{document}